\documentclass[a4paper]{jpconf}
\usepackage{graphicx}
\newcommand{\met}{\mbox{$E\!\!\!\!/_{T}$}}
\newcommand{\fb}{\ensuremath{\rm fb^{-1}}}

\newcommand{\et}{\ensuremath{E_T}}
\newcommand{\etadet}{\ensuremath{\eta_\mathrm{det}}}
\newcommand{\gev}{\ensuremath{\rm GeV}}

\newcommand{\gevcsq}{\ensuremath{{\rm GeV}/c^{2}}}
\newcommand{\gravitino}{\ensuremath{\widetilde{G}}}
\newcommand{\NONE}{\mbox{$\widetilde{\chi}_1^0$}}
\newcommand{\NTWO}{\mbox{$\widetilde{\chi}_2^0$}}
\newcommand{\CONE}{\mbox{$\widetilde{\chi}_1^{\pm}$}}
\newcommand{\CONEP}{\mbox{$\widetilde{\chi}_1^{+}$}}
\newcommand{\CONEM}{\mbox{$\widetilde{\chi}_1^{-}$}}

\begin{document}
\title{Search for New Physics with Photons at the Tevatron}

\author{Shin-Shan Yu on behalf of the CDF and D0 collaborations}

\address{Fermilab, MS~318, P.O.~Box~500, Batavia, IL~60510, USA}

\ead{eikoyu@fnal.gov}

\begin{abstract}
We report on a search for compositeness in $ee\gamma$ events and a 
search for gauge-mediated supersymmetry in $\gamma\gamma\met$ events. We 
also report on two signature-based searches for anomalous production of 
$\gamma\gamma + X$ (where $X=e, \mu, \gamma, \met$) and $\gamma\ell + X$ 
(where $X=\met, \ell, \gamma$). The analyses are based on 
0.9--1.2~\fb\ of data from $p\bar{p}$ collisions at $\sqrt{s}=1.96$ 
TeV collected with the CDF and D0 detectors at the Fermilab Tevatron. 
No significant excess of data over the predicted background has been 
observed. 
\end{abstract}

\section{Introduction}
Physics beyond standard model (SM) predicts a rich spectrum of inclusive 
photon final states with additional detectable objects, such as leptons, 
heavy-flavor jets, and missing transverse energy (\met). Examples of 
the extension of SM include technicolor, large extra dimensions, 
4$^\mathrm{th}$ generation, compositeness, and supersymmetry (SUSY): 
gravity-mediated (mSUGRA) and gauge-mediated (GMSB) supersymmetry breaking. 
The D0 collaboration has performed searches for compositeness in $ee\gamma$~\cite{estarnote} and GMSB in $\gamma\gamma\met$~\cite{gmsbnote} signatures, 
while the CDF collaboration chose to conduct model-independent 
searches for anomalous production 
of $\gamma\gamma + X$ (where $X=e, \mu, \gamma, \met$)~\cite{ggXnote} and 
$\gamma\ell + X$ (where $X=\met, \ell, \gamma$)~\cite{lgXnote}. 
The following sections present results of 
the searches mentioned above. Data for these analyses were collected from 
February 2002 to February 2006. The CDF and D0 detectors are described in 
detail in~\cite{CDF} and~\cite{D0}. 
 
\section{D0 Search for excited electrons in the $ee\gamma$ channel}
\begin{figure}[h]
\begin{minipage}{12pc}
\includegraphics[width=12pc, height=12pc]{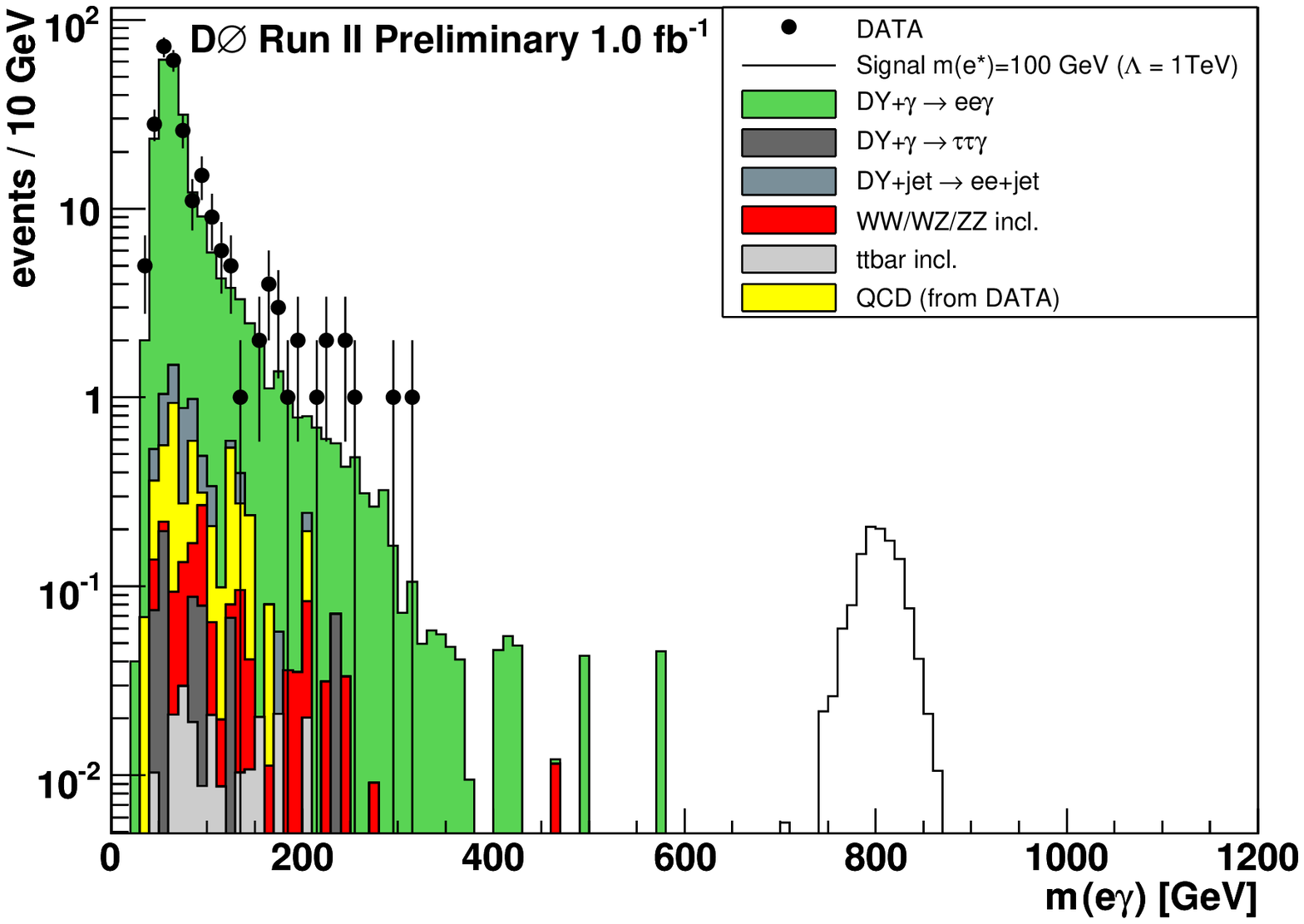}
\caption{\label{fig:estar}The $M_{e\gamma}$ distribution from the D0 data and
 the expected background for mass of $M_{e^*}$=800~\gevcsq. }
\end{minipage}\hspace{1pc}%
\begin{minipage}{12pc}
\includegraphics[width=12pc, height=12pc]{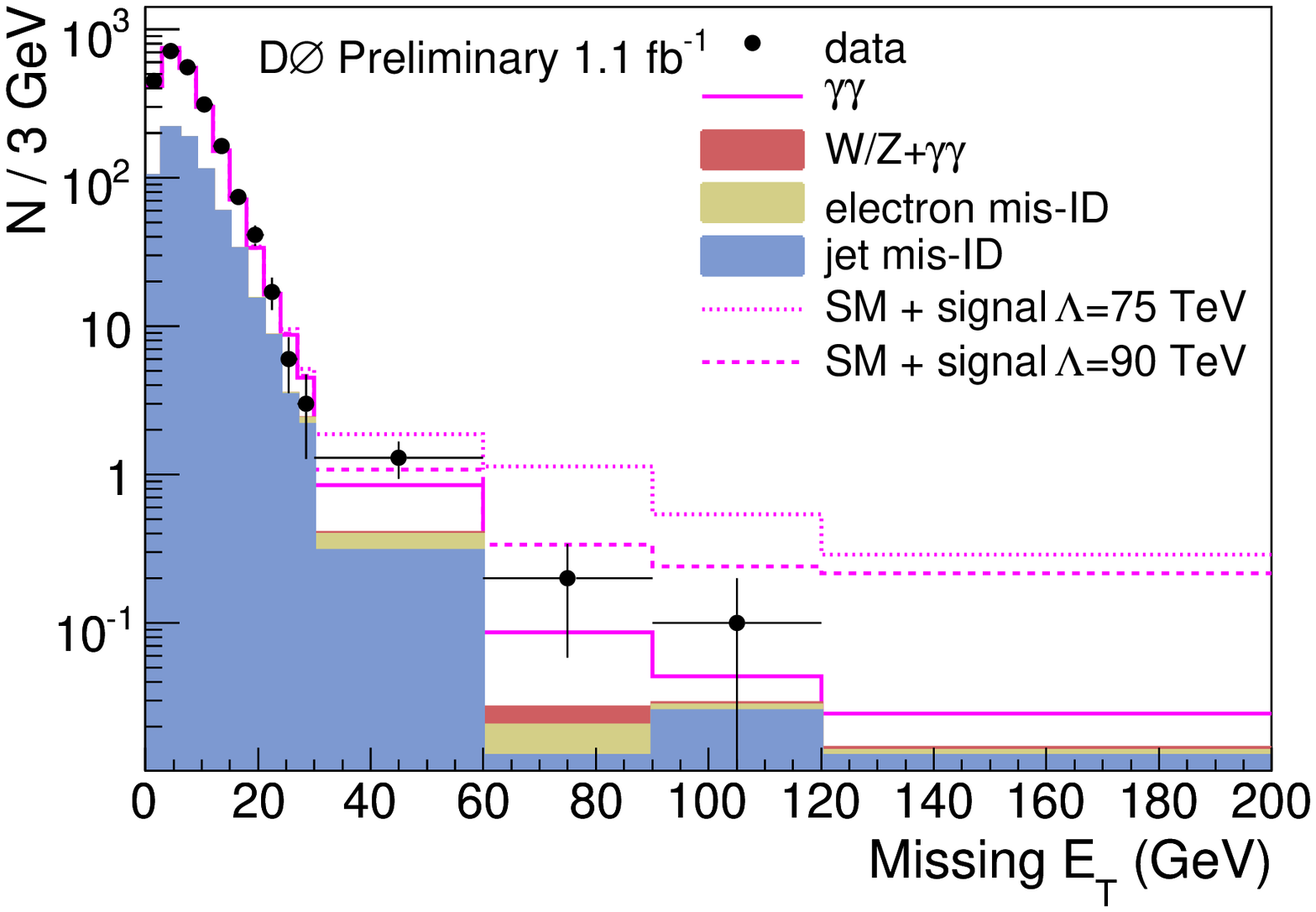}
\caption{\label{fig:gmsb}The \met\ distribution from the D0 
$\gamma\gamma$ data, the expected background, and the GMSB SUSY signal.}
\end{minipage}\hspace{1pc}%
\begin{minipage}{12pc}
\includegraphics[width=12pc, height=12pc]{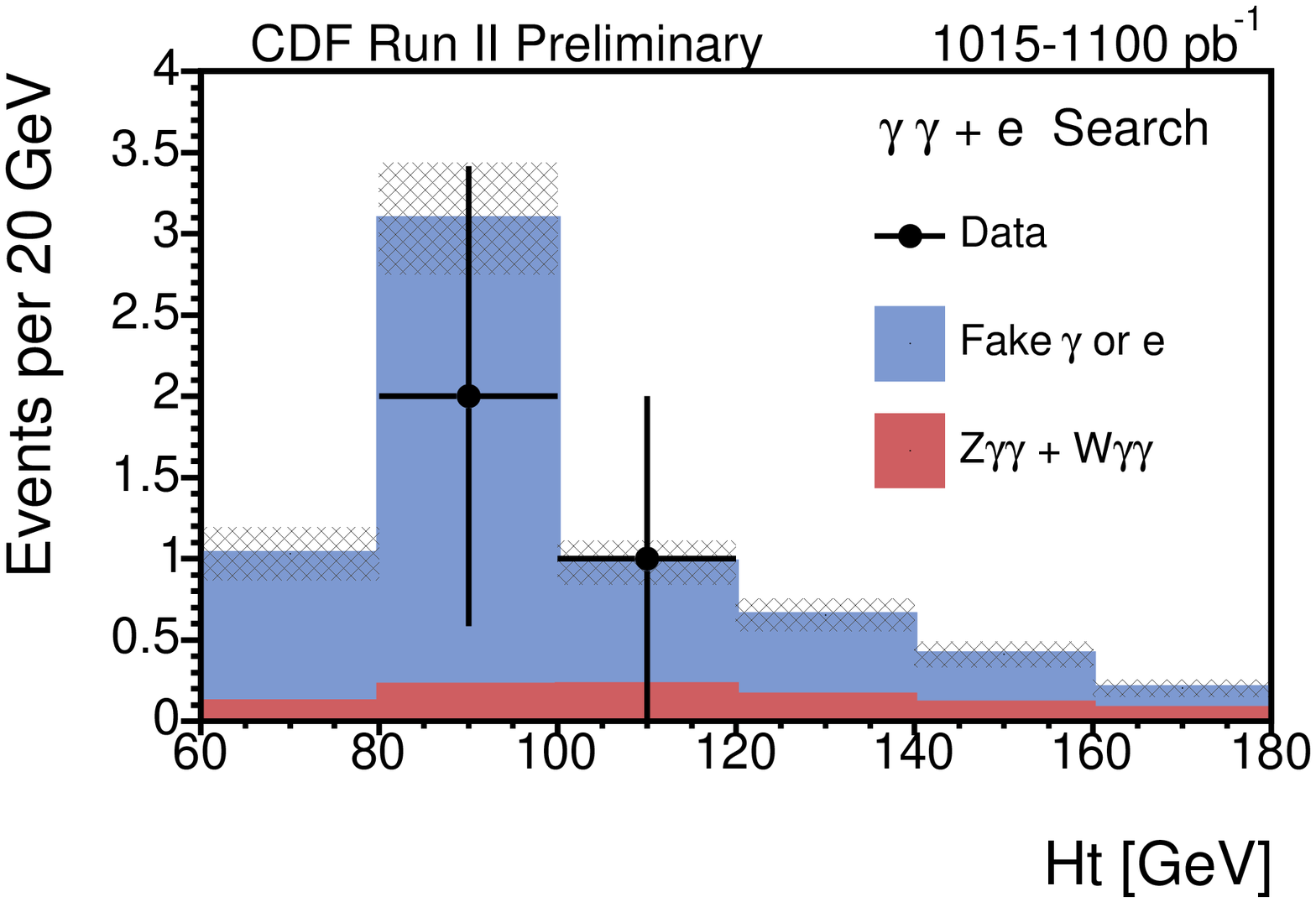}
\caption{\label{fig:gge}The $H_t$ distribution~\cite{ht} of CDF 
$\gamma\gamma e$ events and the expected background.}
\end{minipage} 
\end{figure}

The compositeness model postulates that SM quarks and 
leptons are composed of yet-to-be-observed scalar and spin-1/2 elementary 
particles. The substructure within quarks 
and leptons implies a large spectrum of excited states, including excited 
electrons~\cite{Terazawa:1981eg}. In this search, the excited electrons ($e^*$) are assumed to be 
produced via contact interactions $p\bar{p}\rightarrow ee^*$, with subsequent 
decay of $e^*$ into $e\gamma$ and resulting in a $ee\gamma$ final state. 
We select events containing two isolated electrons with transverse 
energy (\et) greater than 25 and 15~\gev, and one isolated 
photon with \et\ greater than 15~\gev, all within the detector 
pseudorapidity~\cite{eta} range of $|\etadet| < 1.1$ or 
$1.5 < |\etadet| < 2.5$. 
We further optimize the requirements on $M_{e\gamma}$ and $\Delta R(e,\gamma)$
~\cite{deltaR} as a function of the mass of the excited electron ($M_{e^*}$). 
The dominant background after all requirements is the 
Drell-Yan process $Z/\gamma^*\rightarrow ee$ where an additional photon is 
radiated (see Figure~\ref{fig:estar}). Without any excess observed in 1~\fb\ 
of data, we set a 95$\%$ confidence level (CL) lower limit on the $M_{e^*}$ 
as a function of the compositeness scale $\Lambda$. For $\Lambda=1$~TeV, the 
lower limit is found to be 756~\gevcsq. 

\section{D0 Search for gauge-mediated supersymmetry breaking in the $\gamma\gamma\met$ events}
 The lightest SUSY particle in GMSB is the gravitino (\gravitino) with a 
mass of a few keV. 
Assuming $R$-parity conservation, pair production of massive SUSY particles, 
such as \NTWO\CONE\ or \CONEP\CONEM, results in a final state with two photons
 and large \met\ due to the escape of the \gravitino\ from the detector. This 
analysis considers a minimal GMSB model (Snowmass Slope SPS 8~\cite{Martin:2002zb}) with only one 
dimensioned parameter $\Lambda$ that determines the effective scale of SUSY 
breaking. We look for two isolated central photons ($|\etadet| < 1.1$) with 
\et\ greater than 25~\gev\ each. 
The \met\ is determined from energy deposited in the calorimeter for 
$|\etadet| < $ 4.0. 
We compare the observed \met\ distribution with that from the expected. 
The dominant expected background arises from mis-measured \met\ which we 
model with two templates from: 1. $Z\rightarrow ee$ 
events (two real electromagnetic objects), 2. multijet sample, by reversing 
electron identification requirements (mis-identified jet). 
We find no significant discrepancy in the \met\ distribution in 1.1~\fb\ of 
data (see Figure~\ref{fig:gmsb}), and obtain the following 95$\%$ CL limits: 
$\Lambda>$ 88.5~TeV, $M_{\NONE}>$ 120~\gevcsq, and $M_{\CONEP}>$ 220~\gevcsq.

\section{CDF Search for anomalous production of $\gamma\gamma+X$ ($X=e, \mu, \gamma, \met$)}
\begin{figure}[h]
\begin{minipage}{12pc}
\includegraphics[width=12pc, height=12pc]{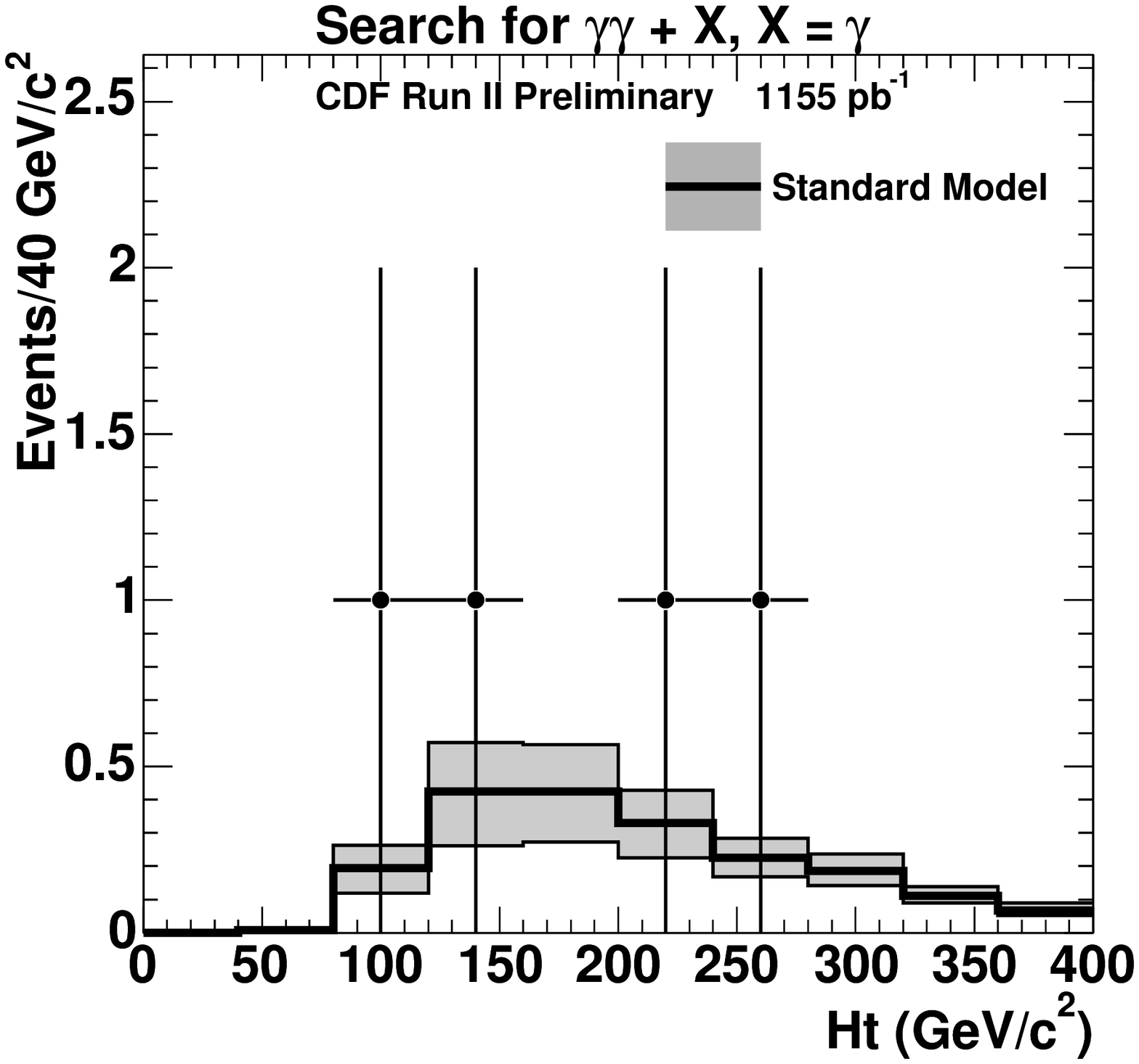}
\caption{\label{fig:ggg}The $H_t$ distribution~\cite{ht} of CDF 
$\gamma\gamma\gamma$ events and the expected background.}
\end{minipage}\hspace{1pc}%
\begin{minipage}{12pc}
\includegraphics[width=12pc, height=12pc]{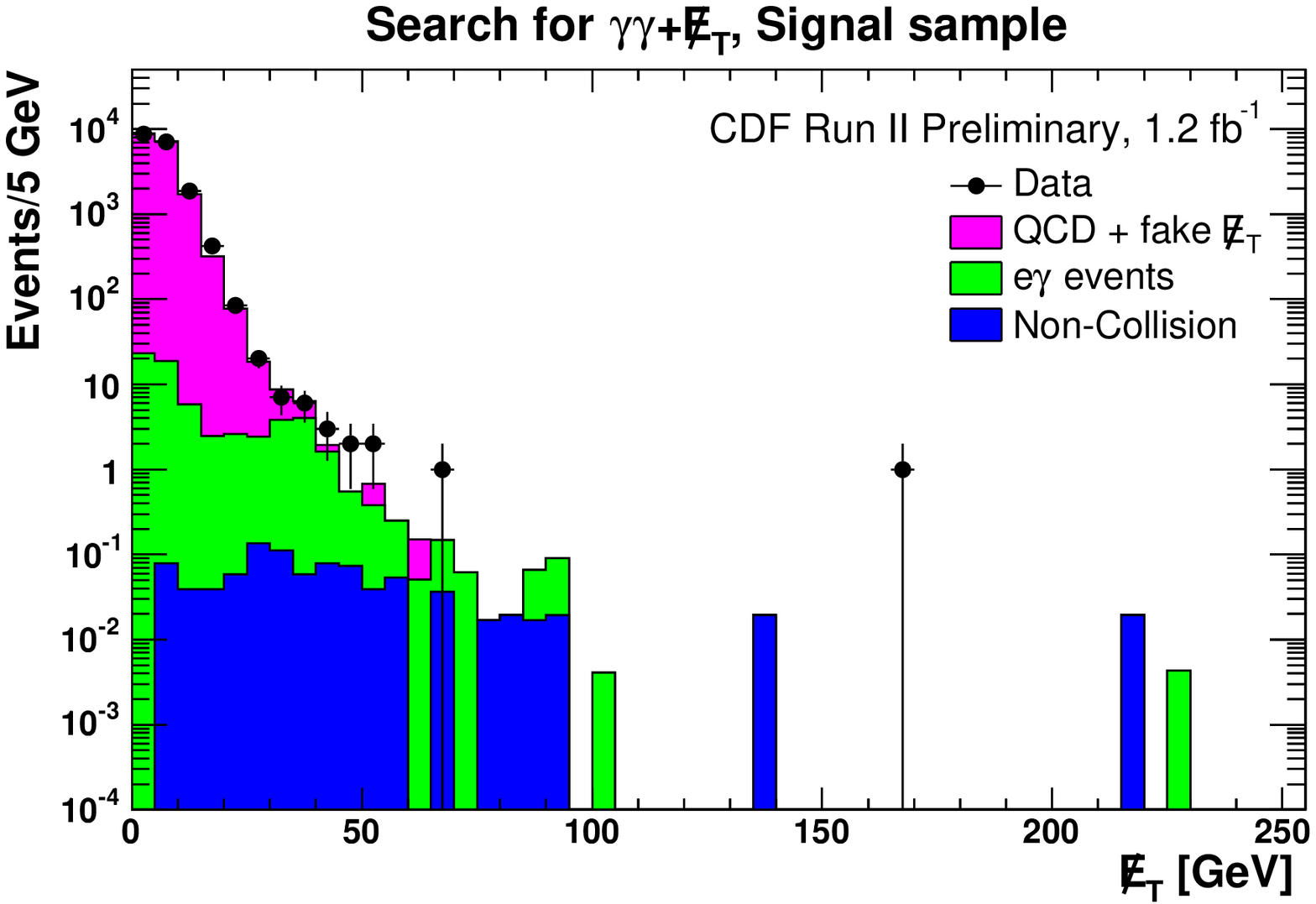}
\caption{\label{fig:ggmet}The \met\ distribution of CDF $\gamma\gamma\met$ 
events and the expected background.}
\end{minipage}\hspace{1pc}%
\begin{minipage}{12pc}
\includegraphics[width=12pc, height=12pc]{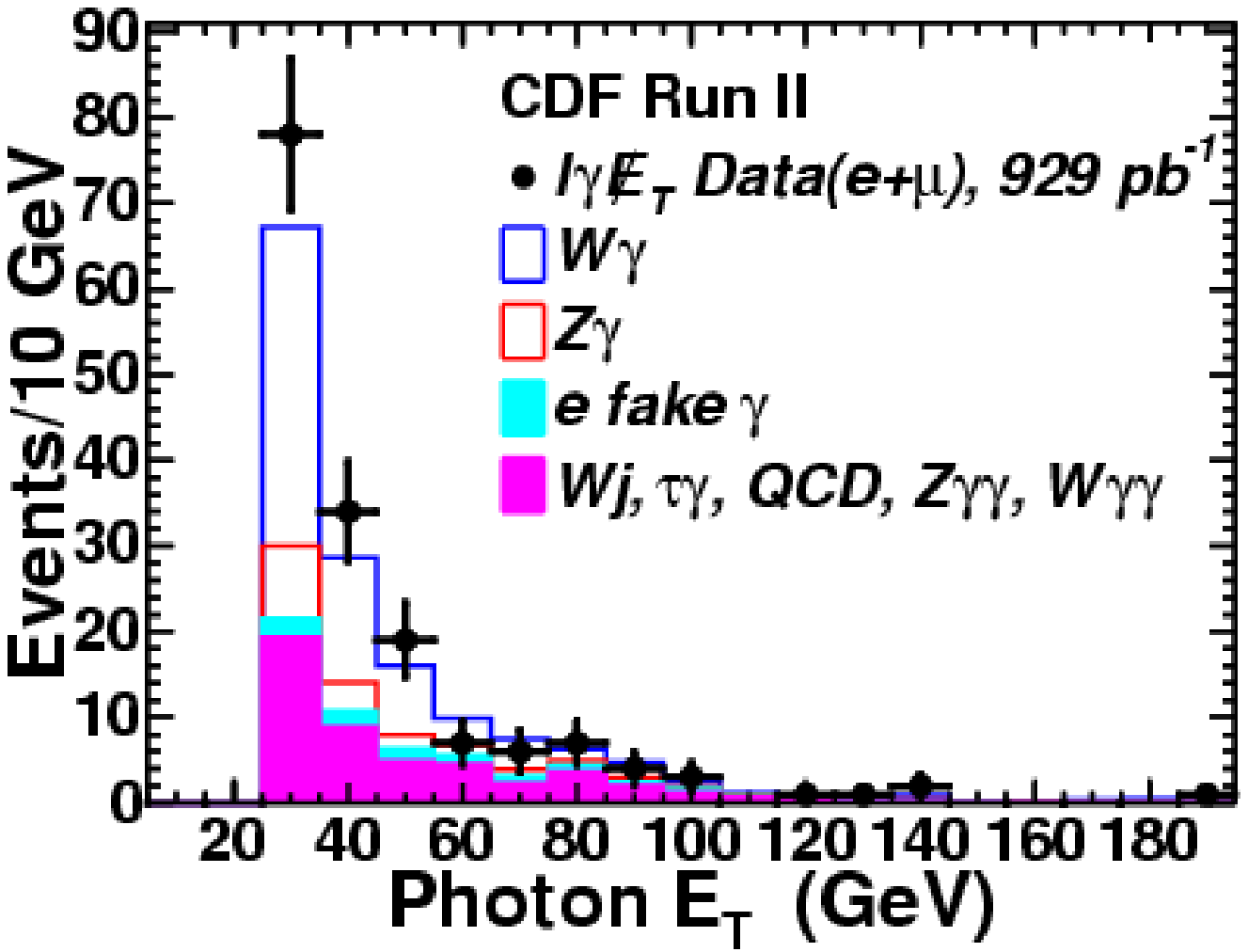}
\caption{\label{fig:lgx}The central photon \et\ distribution of CDF $\gamma\ell\met$ events and the expected background.}
\end{minipage} 
\end{figure}

 We first select events containing two isolated central photons 
($|\etadet| < 1.1$) with \et\ greater than 13~\gev\ each. Then, we look 
for additional $e$, $\mu$, $\gamma$ or large \met\ in the diphoton sample. 
In the $\gamma\gamma e$ and $\gamma\gamma\mu$ final states, we require the 
electron ($|\etadet| < 1.1$ or $1.2 < |\etadet| < 2.8$)
or the muon ($|\etadet| < 1.0$) to have \et\ greater than 20~\gev\ and 
be isolated. The dominant background in the electron channel is 
$Z\gamma$ with an electron from $Z$ mis-identified as a photon, while the 
dominant background in the muon channel is SM $Z\gamma\gamma$ production 
with a muon from $Z$ not reconstructed. In the tri-photon search, 
we require the third photon to satisfy the same 
requirements as those for the other two photons. Here, the two leading 
backgrounds are from SM tri-photon production and multi-jets which are 
mis-identified as photons. 
For the $\gamma\gamma\met$ search, the \met\ is determined from energy 
deposited in the calorimeter for $|\etadet| < $ 3.6. 
We model the mis-measured \met\ distribution by studying: 1. the 
resolution of unclustered energy with zero-jet events in the $Z\rightarrow ee$
 and fake photon data, 2. the resolution of jet energy
 with the dijet data. 
The expected \met\ distribution is a sum of 
mis-measured \met, non-collision background (cosmics, beam halo, photo-tube spikes), and inclusive $W\gamma$ events where the electron from $W$ is 
mis-identified as a photon. 
In all four $\gamma\gamma+X$ searches, we observe no significant excess 
in 1.0--1.2~\fb\ of data. Selected kinematic distributions are 
shown in Figures~\ref{fig:gge}--\ref{fig:ggmet}.

\section{CDF Search for anomalous production of $\gamma\ell + X$ 
($X=\met, \ell, \gamma$; $\ell=e, \mu$)}
We select events which contain one isolated central lepton 
($e$,$\mu$) and one isolated central photon with \et\ greater than 
25~\gev\ each. 
In those events we find additional objects: \met\ greater than 25~\gev, 
a central photon with \et\ greater than 25~\gev, a central 
lepton with \et\ greater than 20~\gev, or a forward electron with \et\ 
greater than 15~\gev. Backgrounds from $W\gamma$, $Z\gamma$, $W\gamma\gamma$, 
$Z\gamma\gamma$ processes are estimated with Madgraph~\cite{madgraph}. 
Jets mis-identified as photons are estimated by fitting 
the calorimetric isolation distribution~\cite{iso} in our data to a signal 
isolation template from $Z\rightarrow ee$ events plus a linear background. 
We observe no excess in all channels in 0.9~\fb\ of data (see 
Figure~\ref{fig:lgx}).

\section{Conclusion}
The CDF and D0 collaborations have carried out both model-driven and 
signature-based searches for new physics in the photon final states. There 
are no hints of new physics in 0.9--1.2~\fb\ of data. However, a factor 
of 4-8 times more data are expected by the end of Run II. In addition, we have 
not fully explored all possible photon final states. Photon 
identification requirements may be improved in the near future, too. We 
will keep searching.

\section*{References}
\begin{thebibliography}{100}
%
\bibitem{estarnote}
{\em http://www-d0.fnal.gov/Run2Physics/WWW/results/prelim/NP/N51/N51.pdf}

\bibitem{gmsbnote}
{\em http://www-d0.fnal.gov/Run2Physics/WWW/results/prelim/NP/N54/N54.pdf}

\bibitem{ggXnote}
{\em http://www-cdf.fnal.gov/physics/exotic/r2a/20060908.diphotonPlusX/pub.ps}

\bibitem{lgXnote}
  A.~Abulencia {\it et al.} (CDF Collaboration),
  Phys.\ Rev.\  D {\bf 75}, 112001 (2007).


\bibitem{CDF}
     D.~Acosta {\it et al.} (CDF Collaboration),
    Phys.\ Rev.\ D {\bf 71}, 032001 (2005).

\bibitem{D0}
  V.~M.~Abazov {\it et al.}  (D0 Collaboration),
  Nucl.\ Instrum.\ Meth.\  A {\bf 565}, 463 (2006).

\bibitem{Terazawa:1981eg}
  U.~Baur, M.~Spira, and P.~M.~Zerwas,
  Phys.\ Rev.\  D {\bf 42}, 815 (1990).

\bibitem{eta}
The pseudorapidity, $\etadet$, is defined as $-\ln\tan(\theta/2)$, where 
$\theta$ is the polar angle to the detector origin.

\bibitem{deltaR}
$\Delta R(e,\gamma)\equiv\sqrt{\Delta\phi(e,\gamma)^2 + 
\Delta \eta(e,\gamma)^2}$. Here, $e$ and $\gamma$ are the daughters 
of $e^*$ candidate. 


\bibitem{Martin:2002zb}
  S.~P.~Martin, S.~Moretti, J.~m.~Qian, and G.~W.~Wilson,
in {\it Proc. of the APS/DPF/DPB Summer Study on the Future of Particle Physics (Snowmass 2001), 
pp P346 } ed. N.~Graf.


\bibitem{madgraph}
  F.~Maltoni and T.~Stelzer,
  ``Madgraph'',
  Comput.\ Phys.\ Commum. {\bf 357}, 81 (1994).

\bibitem{iso}
 The calorimetric isolation is defined as: the transverse energy deposited in 
the calorimeter in a cone of $\Delta R\leq0.4$ around the photon cluster 
centroid after subtracting the photon transverse energy. 

\bibitem{ht}
The scalar sum \et\ of all identified objects (\met, $\gamma$, $e$ $\mu$, jet). 

\end{thebibliography}
\end{document}